# Estimating confidence regions of common measures of (baseline, treatment effect) on dichotomous outcome of a population

Li Yin[1] and Xiaoqin Wang[2*]

[1]Department of Medical Epidemiology and Biostatistics, Karolinska Institute, Box 281, SE-171 77, Stockholm, Sweden

[2]Department of Electronics, Mathematics and Natural Sciences, University of Gävle, SE-801 76, Gävle, Sweden (Email: xwg@hig.se).

[*]Corresponding author

**Abstract**

In this article we estimate confidence regions of the common measures of (baseline, treatment effect) in observational studies, where the measure of baseline is baseline risk or baseline odds while the measure of treatment effect is odds ratio, risk difference, risk ratio or attributable fraction, and where confounding is controlled in estimation of both baseline and treatment effect. To avoid high complexity of the normal approximation method and the parametric or non-parametric bootstrap method, we obtain confidence regions for measures of (baseline, treatment effect) by generating approximate distributions of the ML estimates of these measures based on one logistic model.

## 1 Introduction

Suppose that one conducts a randomized trial to investigate the effect of a dichotomous treatment $z$ on a dichotomous outcome $y$ of certain population, where $z = 0, 1$ indicate the active respective control treatments while $y = 0, 1$ indicate positive respective negative outcomes. With a sufficiently large sample, covariates are essentially unassociated with treatments $z$ and thus are not confounders. Let $\mathrm{R}_z = \mathrm{pr}(y = 1 \mid z)$ be the risk of $y = 1$ given $z$. Then $\mathrm{R}_z$ is marginal with respect to covariates and thus conditional on treatment $z$ only, so $\mathrm{R}_z$ is also called marginal risk. One can use $\mathrm{R}_z$ to obtain the measures of baseline and treatment effect which are called marginal measures. The comment measures of baseline are the baseline risk $\mathrm{R}_0 = \mathrm{pr}(y = 1 \mid z = 0)$ and the baseline odds $\mathrm{O}_0 = \mathrm{R}_0/(1 - \mathrm{R}_0)$.

The common measures of treatment effect are the odds ratio

$$\mathrm{OR} = \frac{\mathrm{O}_1}{\mathrm{O}_0}$$

where $\mathrm{O}_z = \mathrm{R}_z/(1 - \mathrm{R}_z)$ is the odds under treatment $z$, and the risk difference

$$\mathrm{RD} = \mathrm{R}_1 - \mathrm{R}_0,$$

the risk ratio

$$\mathrm{RR} = \frac{\mathrm{R}_1}{\mathrm{R}_0},$$

and the attributable fraction

$$\mathrm{AF} = 1 - \frac{\mathrm{R}_0}{\mathrm{R}_1}.$$

When presenting treatment effect, one also presents baseline in order to learn significance of the treatment effect to the population. For instance, RD = $-$ 0.5 % has rather different epidemiologic meanings when $\mathrm{R}_0$ is 1 % versus 50 %, so one should present both RD and $\mathrm{R}_0$.

A similar situation occurs in which RR = 0.5 when $R_0$ is 1 % versus 50 %. The common measures of (baseline, treatment effect) are $(O_0, \mathrm{OR})$ $(R_0, \mathrm{RD})$, $(R_0, \mathrm{RR})$ and $(R_0, \mathrm{AF})$. All these measures have causal interpretation in the framework of the Rubin causal model and reflect different aspects of the same underlying (baseline, treatment effect) (Rosenbaum and Rubin, 1983; Rosenbaum, 1995; Rubin et al., 2009).

Now suppose that one conducts an observational study to investigate the effect of the treatment *z* on the outcome *y* of the population above and wishes to describe (baseline, treatment effect) by the same measures $(O_0, \mathrm{OR})$ $(R_0, \mathrm{RD})$, $(R_0, \mathrm{RR})$ and $(R_0, \mathrm{AF})$ as in the randomized trial. However, in the observational study, covariates may be associated with both *y* and *z* and therefore are confounders. In this case, one needs to use the conditional risk $\mathrm{pr}(y = 1|z, \boldsymbol{x})$ given not only *z* but also $\boldsymbol{x}$, instead of the marginal risk $\mathrm{pr}(y = 1|z)$, to obtain these measures. As a result, it is highly difficult to estimate these measures as described below.

For illustration, consider the case in which there is only one covariate *x*. To estimate the measures $(O_0, \mathrm{OR})$ $(R_0, \mathrm{RD})$, $(R_0, \mathrm{RR})$ and $(R_0, \mathrm{AF})$ through $\mathrm{pr}(y = 1|z, \boldsymbol{x})$, the most common model is logistic model, for instance,

$$\mathrm{Log}\left\{\frac{\mathrm{pr}(y = 1 \mid z, \boldsymbol{x})}{1 - \mathrm{pr}(y = 1 \mid z, \boldsymbol{x})}\right\} = \alpha + \beta\, z + \theta x + \gamma(zx)$$

Here exp(α) is the baseline odds in stratum $z = 0, x = 0$; exp(β) is the conditional odds ratio describing the treatment effect in stratum $x = 0$. Noticeably, the measures $R_0$, $O_0$, OR, RD, RR and AF as parameters of interest are not equal to the model parameters or their exponentials and therefore cannot be estimated as parameters of this model. Even if $\gamma = 0$, i.e. there is no product term between *z* and *x*, the conditional odds ratio exp(β) is not equal to the marginal odds ratio OR; see rich literatures about non-collapsibility of odds ratios (e.g. Gail, 1984; Greenland et al., 1999; Lee and Nelder, 2004; Austin et al., 2007; Austin, 2007). Also see literatures about convertibility and non-convertibility from exp(β) to RR (Greenland, 1987; Zhang and Yu, 1998; McNutt et al., 2003; Greenland, 2004a); notice literatures about linear or log-linear models in estimation of RD or RR (Bieler et al., 2010; Spiegelman and Hertzmark, 2005; Cheung, 2007; Zou, 2004).

On the other hand, one can always express $R_0$, $O_0$, OR, RD, RR and AP in terms of the parameters of a logistic model. The ML estimates of $R_0$, $O_0$, OR, RD, RR and AP can be obtained from those of the model parameters. The usual methods of obtaining the interval estimate of treatment effect are the normal approximation method and the bootstrap method (Lane and Nelder, 1982; Graubard and Korn, 1999; Greenland, 2004a; Greenland, 2004b; Austin, 2010; Gehrmanna et al., 2010). In the normal approximation method, one derives approximate variance of the ML estimate of a measure of treatment effect and then uses the variance to obtain normal approximation confidence interval for the measure. However, the derivation of the variance is tedious and the expressions of the variance are different for different logistic models containing different products between $z$ and $\boldsymbol{x}$. In the bootstrap method, one generates bootstrap samples and then uses the bootstrap samples to obtain bootstrap distribution and thus the confidence interval of a measure of treatment effect. However, it is highly difficult to correct finite-sample bias due to the bootstrap sampling, which typically occurs with the asymmetric – i.e. screwed -- distribution of the ML estimate as is the case for dichotomous outcome (Davison and Hinkley, 1997; Carpenter and Bithell, 2000; Greenland, 2004b).

In this article, we use only one logistic model to obtain the confidence regions for all the measures $(\mathrm{O}_0, \mathrm{OR})$, $(\mathrm{R}_0, \mathrm{RD})$, $(\mathrm{R}_0, \mathrm{RR})$ and $(\mathrm{R}_0, \mathrm{AP})$ of (baseline, treatment effect) in observational studies. Instead of deriving covariance matrix of the two-dimensional ML estimate or correcting the finite-sample bias of the two-dimensional bootstrap sampling, we use approximate normal distribution of the ML estimate of the logistic model parameters to generate approximate distribution of the ML estimate of a measure of (baseline, treatment effect) and then use the obtained distribution to calculate confidence region for this measure.

We are going to present the method through an observational study embedded in a randomized trial, in which we investigated the effect of high versus low doses of a therapy on eradication of *Helicobacter pylori* among Vietnamese children.

## 2 Eradication of *Helicobacter pylori* among Vietnamese children

In a randomized trial, researchers studied two triple therapies – (lansoprazole, amoxicillin, metronidazole) and (lansoprazole, amoxicillin, clarithromycin) – for their abilities to eradicate *Helicobacter pylori* among Vietnamese children (Nguyen et al., 2008). From several children hospitals in Hanoi, restricting body weight to a range between 13 kg and 45 kg, a sample was collected between May 2005 and January 2006. In a substudy, researchers focused on one treatment arm of the triple therapy (lansoprazole, amoxicillin, metronidazole) and analyzed the treatment effect of high versus low doses of the therapy on eradication of *Helicobacter pylori*. The treatment arm comprised 109 patients.

The therapy eradicated *Helicobacter pylori* through systemic circulation, so the researchers assigned the therapy to the children according to their body weights. According to the pediatric procedure, children with 13 kg < body weight < 23 kg received the therapy once daily and those with 23 kg ≤ weight < 45 kg twice daily. Of medical relevance was dose in unit body weight. Following the earlier research on the same data, we categorized the children into receiving high versus low doses of the therapy by taking the middle point of each weight category as the threshold (Nguyen et al., 2008). Hence children with 13 kg < body weight ≤18 kg in the first category and 23 kg ≤ body weight < 34 kg in the second category were considered as receiving the high dose ($z$ = 1) of the therapy whereas those with 18 kg < body weight < 23 kg in the first category and 34 kg ≤ body weight < 45 kg in the second category as receiving the low dose ($z$ = 0).

The outcome was successful ($y$ = 1) versus unsuccessful ($y$ = 0) eradication of *Helicobacter pylori*.

In the context of this substudy, body weight was synonymous to treatment $z$ and thus was not a covariate. The covariate age ($x_1$) was associated with body weight and thus with the assignment of treatments $z$. Furthermore, age was known to have effect on the outcome $y$ and thus was a confounder. Therefore this substudy had the nature of an observational study. Other possible confounding covariates were documented, which were gender ($x_2$), geographic area ($x_3$) and antibiotic resistance to metronidazole ($x_4$). Age was categorized into younger ($x_1$ = 1) than 9 years *versus* older ($x_1$ = 0). Let $x_2$ = 1 indicate female and $x_2$ = 0 male of the gender. Geographic area was categorized into rural ($x_3$ = 1) *versus* urban ($x_3$ = 0). Antibiotic resistance to metronidazole was categorized into sensitive ($x_4$ = 1) *versus* resistant ($x_4$ = 0). Let $\boldsymbol{x}$ = ($x_1$, $x_2$, $x_3$, $x_4$) be the set of the documented covariates. The data of the substudy is given in Table 1.

The conditional risk $\mathrm{pr}(y = 1 \mid z, \boldsymbol{x})$ of $y = 1$ in stratum $(z, \boldsymbol{x})$ is modeled by logistic model. By the likelihood ratio-based significance testing of the model parameters, we obtain

$$\mathrm{Log}\{\frac{\mathrm{pr}(y = 1 \mid z, \boldsymbol{x})}{1 - \mathrm{pr}(y = 1 \mid z, \boldsymbol{x})}\} =$$

$$\alpha + \beta\, z + \theta_1 x_1 + \theta_2 x_2 + \theta_3 x_3 + \theta_4 x_4 + \theta_5 (z * x_2) + \theta_6 (z * x_3). \qquad (1)$$

In addition to the main effect terms for the treatment *z* and the covariates $\boldsymbol{x}$, we include in this model two product terms: one for the treatment--gender ($z * x_2$) and one for the treatment--geographic area ($z * x_3$), because of the small p-values, 0.10 and 0.11, for the significance test of $\theta_5 = 0$ and $\theta_6 = 0$ respectively. Let $\pi = (\alpha,\ \beta,\ \theta_1,\ \theta_2\ ,\ \theta_3,\ \theta_4,\ \theta_5,\ \theta_6)$ be the set of all the model parameters. The ML estimate $\hat{\pi} = (\hat{\alpha},\ \hat{\beta},\ \hat{\theta}_1,\ \hat{\theta}_2,\ \hat{\theta}_3,\ \hat{\theta}_4,\ \hat{\theta}_5,\ \hat{\theta}_6)$ and its approximate covariance matrix $\hat{\Sigma}$ (i.e. the inverse of the observed information) are presented in Table 2.

## 3 Confidence regions of (baseline, treatment effect)

### 3.1 Measures of (baseline, treatment effect)

Here we use the conditional risk $\mathrm{pr}(y = 1 \mid z, \boldsymbol{x})$ to obtain the measures $(\mathrm{O_0, OR})$, $(\mathrm{R_0, RD})$, $(\mathrm{R_0, RR})$ and $(\mathrm{R_0, AF})$ of (baseline, treatment effect) introduced in the introduction. Let $y_z$ be the potential outcome of each patient in the population under treatment *z = 0, 1*. Denote the risk $\mathrm{pr}(y_z = 1)$ of the potential outcome $y_z = 1$ by $\mathrm{R}_z$. Then $\mathrm{R_0}$ is the baseline risk, i.e. the risk of $y_z = 1$ in the population under *z* = 0.

Because it is only possible to observe potential outcome of a patient under either *z* = 0 or *z* = 1, certain assumption is needed to allow for estimation of $\mathrm{R}_z$ (Greenland et al., 1999; Greenland and Robins, 1986; Rosenbaum and Rubin, 1983; Rosenbaum, 1995; Rubin et al., 2009). In the medical context of this study, it is reasonable to assume that there is no other confounding covariate than the documented covariates $\boldsymbol{x}$ (Nguyen et al., 2008). The assumption implies

$$\mathrm{pr}(y_z = 1|\boldsymbol{x}) = \mathrm{pr}(y = 1 \mid z, \boldsymbol{x})$$

which states that the patients in stratum $\boldsymbol{x}$, who have not received treatment $z$, would have the same risk $\mathrm{pr}(y = 1 \mid z, \boldsymbol{x})$ as those who have received treatment $z$, if they had received treatment $z$. Therefore we can express the risk $\mathrm{R}_z = \mathrm{pr}(y_z = 1)$ of the potential outcome $y_z = 1$ by

$$\mathrm{R}_z = \sum_{\boldsymbol{x}} \mathrm{pr}(y_z = 1|\boldsymbol{x})\mathrm{pr}(\boldsymbol{x}) = \sum_{\boldsymbol{x}} \mathrm{pr}(y = 1|z, \boldsymbol{x})\mathrm{pr}(\boldsymbol{x}), \tag{2}$$

and in particular, the baseline risk $\mathrm{R}_0$ by

$$\mathrm{R}_0 = \sum_{\boldsymbol{x}} \mathrm{pr}(y = 1|z = 0, \boldsymbol{x})\mathrm{pr}(\boldsymbol{x}),$$

where $\mathrm{pr}(\boldsymbol{x})$ is the probability of $\boldsymbol{x}$ in the population. Formula (2) implies that we can estimate $\mathrm{R}_z$ or its function by modeling $\mathrm{pr}(y = 1|z, \boldsymbol{x})$ and $\mathrm{pr}(\boldsymbol{x})$ through the observed data.

Now we express the measures of baseline and treatment effect in terms of $\mathrm{pr}(y = 1|z, \boldsymbol{x})$ and $\mathrm{pr}(\boldsymbol{x})$. The odds of the population under treatment $z$ has the form

$$\mathrm{O}_z = \frac{\mathrm{R}_z}{1 - \mathrm{R}_z} = \frac{\sum_{\boldsymbol{x}} \mathrm{pr}(y = 1 \mid z, \boldsymbol{x})\mathrm{pr}(\boldsymbol{x})}{1 - \sum_{\boldsymbol{x}} \mathrm{pr}(y = 1 \mid z, \boldsymbol{x})\mathrm{pr}(\boldsymbol{x})}, \quad z = 0,1 \tag{3}$$

Hence the baseline odds is

$$\mathrm{O}_0 = \frac{\mathrm{R}_0}{1 - \mathrm{R}_0} = \frac{\sum_{\boldsymbol{x}} \mathrm{pr}(y = 1 \mid z = 0, \boldsymbol{x})\mathrm{pr}(\boldsymbol{x})}{1 - \sum_{\boldsymbol{x}} \mathrm{pr}(y = 1 \mid z = 0, \boldsymbol{x})\mathrm{pr}(\boldsymbol{x})}. \tag{4}$$

The odds ratio is

$$\mathrm{OR} = \frac{\mathrm{O}_1}{\mathrm{O}_0}$$

$$= \frac{\sum_{\boldsymbol{x}} \mathrm{pr}(y = 1 \mid z = 1, \boldsymbol{x})\mathrm{pr}(\boldsymbol{x})}{1 - \sum_{\boldsymbol{x}} \mathrm{pr}(y = 1 \mid z = 1, \boldsymbol{x})\mathrm{pr}(\boldsymbol{x})} \Big/ \frac{\sum_{\boldsymbol{x}} \mathrm{pr}(y = 1 \mid z = 0, \boldsymbol{x})\mathrm{pr}(\boldsymbol{x})}{1 - \sum_{\boldsymbol{x}} \mathrm{pr}(y = 1 \mid z = 0, \boldsymbol{x})\mathrm{pr}(\boldsymbol{x})}. \quad (5)$$

The risk difference is

$$\mathrm{RD} = \mathrm{R}_1 - \mathrm{R}_0 = \sum_{\boldsymbol{x}} \{\mathrm{pr}(y = 1|z = 1, \boldsymbol{x}) - \mathrm{pr}(y = 1|z = 0, \boldsymbol{x})\}\mathrm{pr}(\boldsymbol{x}). \quad (6)$$

The risk ratio is

$$\mathrm{RR} = \frac{\mathrm{R}_1}{\mathrm{R}_0} = \frac{\sum_{\boldsymbol{x}} \mathrm{pr}(y = 1 \mid z = 1, \boldsymbol{x})\mathrm{pr}(\boldsymbol{x})}{\sum_{\boldsymbol{x}} \mathrm{pr}(y = 1 \mid z = 0, \boldsymbol{x})\mathrm{pr}(\boldsymbol{x})}. \quad (7)$$

Finally, the attributable fraction is

$$\mathrm{AF}=1 - \frac{\mathrm{R}_0}{\mathrm{R}_1} = 1 - \frac{\sum_{\boldsymbol{x}} \mathrm{pr}(y = 1 \mid z = 0, \boldsymbol{x})\mathrm{pr}(\boldsymbol{x})}{\sum_{\boldsymbol{x}} \mathrm{pr}(y = 1 \mid z = 1, \boldsymbol{x})\mathrm{pr}(\boldsymbol{x})}. \quad (8)$$

In particular, in randomized trial, the treatment $z$ is not associated with the covariates $\boldsymbol{x}$, so that $\mathrm{pr}(\boldsymbol{x}) = \mathrm{pr}(\boldsymbol{x}|z)$. Insertion of this into (2) yields the risk $\mathrm{pr}(y = 1 |z)$ of $y = 1$ in stratum $z$,

$$\mathrm{R}_z = \sum_{\boldsymbol{x}} \mathrm{pr}(y = 1|z, \boldsymbol{x})\mathrm{pr}(\boldsymbol{x}|z) = \mathrm{pr}(\mathrm{y} = 1|\mathrm{z}).$$

Therefore we see that $\mathrm{R}_z$ and thus $\mathrm{R}_0$, $\mathrm{O}_0$, OR, RD, RR and AF defined here are equal to those defined in the introduction for randomized trial. No matter if it is a randomized trial or an observational study, all these measures have causal interpretation in the Rubin causal model (Rosenbaum and Rubin, 1983; Rosenbaum, 1995; Rubin et al., 2009).

## 3.2 Point estimation of (baseline, treatment effect)

From model (1), we obtain

$$\mathrm{pr}(y = 1 \mid z, \boldsymbol{x}) =$$

$$\frac{\exp\{\alpha + \beta\, z + \theta_1 x_1 + \theta_2 x_2 + \theta_3 x_3 + \theta_4 x_4 + \theta_5 (z * x_2) + \theta_6 (z * x_3)\}}{1 + \exp\{\alpha + \beta\, z + \theta_1 x_1 + \theta_2 x_2 + \theta_3 x_3 + \theta_4 x_4 + \theta_5 (z * x_2) + \theta_6 (z * x_3)\}}. \qquad (9)$$

We are going to use (9) to express the measures of baseline and treatment effect in terms of the model parameters $\pi = (\alpha,\ \beta,\ \theta_1,\ \theta_2\ ,\ \theta_3,\ \theta_4,\ \theta_5,\ \theta_6)$. We shall ignore variability of the covariates $\boldsymbol{x}$, so that we replace the probability $\mathrm{pr}(\boldsymbol{x})$ by the proportion $\mathrm{prop}(\boldsymbol{x})$ and treat the proportion as constant.

Inserting $\mathrm{pr}(y = 1 \mid z, \boldsymbol{x})$ of (9) into (2), we obtain $\mathrm{R}_z$ $(z = 0{,}1)$ as a function of $\pi$,

$$\mathrm{R}_z(\pi) = \sum_{\boldsymbol{x}} \mathrm{pr}(y = 1 | z, \boldsymbol{x}) \mathrm{prop}(\boldsymbol{x}). \qquad (10)$$

Inserting $\mathrm{pr}(y = 1 \mid z, \boldsymbol{x})$ of (9) into (3)-(5), we obtain $\mathrm{O}_z$ and OR as functions of $\pi$,

$$\mathrm{O}_z(\pi) = \frac{\mathrm{R}_z(\pi)}{1 - \mathrm{R}_z(\pi)}, \quad z = 0{,}1 \qquad (11)$$

$$\mathrm{OR}(\pi) = \frac{\mathrm{O}_1(\pi)}{\mathrm{O}_0(\pi)} \qquad (12)$$

Similarly, from (6)-(8), we obtain

$$\mathrm{RD}(\pi) = \mathrm{R}_1(\pi) - \mathrm{R}_0(\pi), \qquad (13)$$

$$\mathrm{RR}(\pi) = \frac{\mathrm{R}_1(\pi)}{\mathrm{R}_0(\pi)}, \qquad (14)$$

$$\mathrm{AF}(\pi) = 1 - \frac{\mathrm{R}_0(\pi)}{\mathrm{R}_1(\pi)}. \qquad (15)$$

Replacing $\pi$ in $\mathrm{R}_0(\pi)$, $\mathrm{O}_0(\pi)$, $\mathrm{OR}(\pi)$, $\mathrm{RD}(\pi)$, $\mathrm{RR}(\pi)$ and $\mathrm{AR}(\pi)$ by the ML estimate $\hat{\pi}$ given in Table 2, we obtain the ML estimates $\hat{\mathrm{R}}_0 = \mathrm{R}_0(\hat{\pi})$, $\hat{\mathrm{O}}_0 = \mathrm{O}_0(\hat{\pi})$, $\widehat{\mathrm{OR}} = \mathrm{OR}(\hat{\pi})$, $\widehat{\mathrm{RD}} = \mathrm{RD}(\hat{\pi})$, $\widehat{\mathrm{RR}} = \mathrm{RR}(\hat{\pi})$ and $\widehat{\mathrm{AP}} = \mathrm{AP}(\hat{\pi})$. These estimates are $\hat{\mathrm{R}}_0 = 0.53$, $\hat{\mathrm{O}}_0 = 1.13$, $\widehat{\mathrm{OR}} = 2.11$, $\widehat{\mathrm{RD}} = 0.17$, $\widehat{\mathrm{RR}} = 1.33$, and $\widehat{\mathrm{AP}} = 0.25$, which are also presented in Figures 1-4 and Table 3.

### 3.3 Distributions for ML estimates of (baseline, treatment effect)

Parameters of a logistic model have good asymptotic normality: normal distribution is good approximation to distribution of the ML estimate of the parameters (e.g. Lindsey, 1996). We are going to use the approximate normal distribution of the ML estimate of the model parameters to generate approximate distribution of the ML estimates of measures of (baseline, treatment effect).

Let $p$ be a random variable which follows the normal distribution $\mathrm{N}(\hat{\pi}, \hat{\Sigma})$, namely, $p \sim \mathrm{N}(\hat{\pi}, \hat{\Sigma})$, where $\hat{\pi}$ and $\hat{\Sigma}$ in $\mathrm{N}(\hat{\pi}, \hat{\Sigma})$ are given in Table 2. Replacing $\pi$ in $\{\mathrm{O}_0(\pi), \mathrm{OR}(\pi)\}$ by $p$ gives $\{\mathrm{O}_0(p), \mathrm{OR}(p)\}$. By using the normal distribution of $p$, we generate the distribution of $\{\mathrm{O}_0(p), \mathrm{OR}(p)\}$, which approximates the distribution of $(\hat{\mathrm{O}}_0, \widehat{\mathrm{OR}})$. Using the same method, we also obtain the approximate distributions of $(\hat{\mathrm{R}}_0, \widehat{\mathrm{RD}})$, $(\hat{\mathrm{R}}_0, \widehat{\mathrm{RR}})$ and $(\hat{\mathrm{R}}_0, \widehat{\mathrm{AF}})$ respectively.

### 3.4 Confidence region estimation of (baseline, treatment effect)

In this article, the confidence region of a measure of (baseline, treatment effect) is defined as the smallest area given a confidence level $(1-\alpha)$, i.e. the smallest area in which the ML estimate of the measure has the probability $(1-\alpha)$. Given the distribution of the ML estimate of a measure of (baseline, treatment effect), we can obtain the confidence region of the measure by a variety of methods. Here we are going to obtain confidence regions of $(\mathrm{O}_0, \mathrm{OR})$, $(\mathrm{R}_0, \mathrm{RD})$, $(\mathrm{R}_0, \mathrm{RR})$ and $(\mathrm{R}_0, \mathrm{AP})$ by using convertibility of these measures. As will be seen in Section 4, this method can be easily implemented by using any software that generates normal distribution.

It is clear from (3)-(5) that $(\mathrm{R}_0, \mathrm{R}_1)$ can be converted into $(\mathrm{O}_0, \mathrm{OR})$. On the other hand, from (4), we obtain

$$\mathrm{R}_0 = \frac{\mathrm{O}_0}{1 + \mathrm{O}_0}. \tag{16}$$

Combining with (3) and (5), we obtain

$$\mathrm{R}_1 = \frac{(\mathrm{O}_0)(\mathrm{OR})}{1 + (\mathrm{O}_0)(\mathrm{OR})}. \tag{17}$$

Thus $(\mathrm{O}_0, \mathrm{OR})$ can be converted into $(\mathrm{R}_0, \mathrm{R}_1)$. Therefore $(\mathrm{O}_0, \mathrm{OR})$ and $(\mathrm{R}_0, \mathrm{R}_1)$ can be converted into each other. Furthermore, from (6)-(8), we see that $(\mathrm{R}_0, \mathrm{R}_1)$ can be converted to the measures $(\mathrm{R}_0, \mathrm{RD})$, $(\mathrm{R}_0, \mathrm{RR})$ and $(\mathrm{R}_0, \mathrm{AP})$, and vice versa. As a result, the measures $(\mathrm{R}_0, \mathrm{RD})$, $(\mathrm{R}_0, \mathrm{RR})$, $(\mathrm{O}_0, \mathrm{OR})$ and $(\mathrm{R}_0, \mathrm{AP})$ can be converted, i.e. obtained, from one another.

Similarly, from (12)-(15), we see that the measures of (baseline, treatment effect) as function of the model parameters $\pi$, $\{\mathrm{O}_0(\pi), \mathrm{OR}(\pi)\}$, $\{\mathrm{R}_0(\pi), \mathrm{RD}(\pi)\}$, $\{\mathrm{R}_0(\pi), \mathrm{RR}(\pi)\}$ and $\{\mathrm{R}_0(\pi), \mathrm{AP}(\pi)\}$, are convertible from one another. Because $\widehat{\mathrm{R}}_0 = \mathrm{R}_0(\hat{\pi})$, $\widehat{\mathrm{O}}_0 = \mathrm{O}_0(\hat{\pi})$, $\widehat{\mathrm{OR}} = \mathrm{OR}(\hat{\pi})$, $\widehat{\mathrm{RD}} = \mathrm{RD}(\hat{\pi})$, $\widehat{\mathrm{RR}} = \mathrm{RR}(\hat{\pi})$ and $\widehat{\mathrm{AP}} = \mathrm{AP}(\hat{\pi})$, the ML estimates, $(\widehat{\mathrm{R}}_0, \widehat{\mathrm{RD}})$, $(\widehat{\mathrm{R}}_0, \widehat{\mathrm{RR}})$, $(\widehat{\mathrm{O}}_0, \widehat{\mathrm{OR}})$ and $(\widehat{\mathrm{R}}_0, \widehat{\mathrm{AF}})$, are convertible with one another. Therefore the distributions of these ML estimates are convertible with one another and so are their corresponding confidence regions.

To obtain confidence regions of $(\mathrm{R_0, RD})$, $(\mathrm{R_0, RR})$, $(\mathrm{O_0, OR})$ and $(\mathrm{R_0, AP})$, we consider the logarithm of $(\mathrm{O_0, OR})$ , i.e. $\{\log(\mathrm{O_0}), \log(\mathrm{OR})\}$, as an additional measure of (baseline, treatment effect). Because the logarithm is a monotonic function, $\{\log(\mathrm{O_0}), \log(\mathrm{OR})\}$ is convertible with any other measure of (baseline, treatment effect), and so is its ML estimate $\{\log(\widehat{\mathrm{O}}_0), \log(\widehat{\mathrm{OR}})\}$ with the ML estimate of any other measure, and similarly for the distribution of $\{\log(\widehat{\mathrm{O}}_0), \log(\widehat{\mathrm{OR}})\}$ and its corresponding confidence region.

Because $\log(\mathrm{O_0})$ and $\log(\mathrm{OR})$ range from $-\infty$ to $+\infty$, the ML estimate $\{\log(\widehat{\mathrm{O}}_0), \log(\widehat{\mathrm{OR}})\}$ is fairly normal even for finite sample (e.g. Lindsey, 1996) and thus the corresponding confidence region can be well approximated by an ellipse. Due to the convertibility of confidence regions, we can convert the elliptical confidence region of $\{\log(\mathrm{O_0}), \log(\mathrm{OR})\}$ to non-elliptical confidence regions of $(\mathrm{O_0, OR})$, $(\mathrm{R_0, RD})$, $(\mathrm{R_0, RR})$ and $(\mathrm{R_0, AP})$.

In the next section, we shall describe the procedure of obtaining point estimates and confidence regions of measures of (baseline, treatment effect).

# 4 Procedure for obtaining confidence regions of (baseline, treatment effect)

## 4.1 Approximate distributions for the ML estimates of baseline and treatment effect

First we draw $p$ from the normal distribution $p \sim \mathrm{N}(\hat{\pi}, \hat{\Sigma})$. Second, we replace $\pi$ by the obtained $p$ in formulas (11) and (12) to get $\{\mathrm{O_0}(p), \mathrm{OR}(p)\}$ which approximates $(\widehat{\mathrm{O}}_0, \widehat{\mathrm{OR}})$. We iterate the procedure, 1000 times in this article, to get 1000 pairs of values of $(\widehat{\mathrm{O}}_0, \widehat{\mathrm{OR}})$. All these 1000 pairs form an approximate (joint) distribution of $(\widehat{\mathrm{O}}_0, \widehat{\mathrm{OR}})$. All values of $\widehat{\mathrm{O}}_0$ in those 1000 pairs form an approximate (marginal) distribution of $\widehat{\mathrm{O}}_0$. All values of $\widehat{\mathrm{OR}}$ in those 1000 pairs form an approximate (marginal) distribution of $\widehat{\mathrm{OR}}$. Similarly, we obtain the distributions of other measures of baseline and treatment effect. The approximate distributions for ($\widehat{\mathrm{O}}_0$, $\widehat{\mathrm{OR}}$), $\widehat{\mathrm{O}}_0$ and $\widehat{\mathrm{OR}}$ are presented in Figure 1; those for ($\widehat{\mathrm{R}}_0$, $\widehat{\mathrm{RD}}$), $\widehat{\mathrm{R}}_0$ and $\widehat{\mathrm{RD}}$ in Figure 2; those for ($\widehat{\mathrm{R}}_0$, $\widehat{\mathrm{RR}}$) and $\widehat{\mathrm{RR}}$ in Figure 3; those for ($\widehat{\mathrm{R}}_0$, $\widehat{\mathrm{AF}}$) and $\widehat{\mathrm{AF}}$ in Figure 4.

### 4.2 Confidence regions of (baseline, treatment effect)

By taking logarithm of the 1000 values of $\left(\widehat{\mathrm{O}}_0, \widehat{\mathrm{OR}}\right)$, $\widehat{\mathrm{O}}_0$ and $\widehat{\mathrm{OR}}$, we obtain the distributions of $\{\log(\widehat{\mathrm{O}}_0), \log(\widehat{\mathrm{OR}})\}$, $\log(\widehat{\mathrm{O}}_0)$, $\log(\widehat{\mathrm{OR}})$, which are presented in Figure 5. From the distribution of $\{\log(\widehat{\mathrm{O}}_0), \log(\widehat{\mathrm{OR}})\}$, we calculate the mean and covariance matrix and then construct the following normal distribution

$$\mathrm{N}\left\{\begin{pmatrix} 0.12 \\ 0.68 \end{pmatrix}, \begin{pmatrix} 0.11 & -0.11 \\ -0.11 & 0.21 \end{pmatrix}\right\}.$$

From this normal distribution, we obtain the ellipse formula

$$\{\log(\mathrm{O}_0) - 0.12 \quad \log(\mathrm{OR}) - 0.68\} \begin{pmatrix} 0.11 & -0.11 \\ -0.11 & 0.21 \end{pmatrix} \begin{Bmatrix} \log(\mathrm{O}_0) - 0.12 \\ \log(\mathrm{OR}) - 0.68 \end{Bmatrix} = \chi_2^2(1-\alpha)$$

from which we obtain the $(1-\alpha)$ confidence region of $\{\log(\mathrm{O}_0), \log(\mathrm{OR})\}$, which is plotted in Figure 5. Now we convert this elliptical confidence region into non-elliptical confidence regions for other measures of (baseline, treatment effect) by the following procedure: a) converting $\{\log(\mathrm{O}_0), \log(\mathrm{OR})\}$ to $(\mathrm{O}_0, \mathrm{OR})$, b) using formulas (16) and (17) to obtain $(\mathrm{R}_0, \mathrm{R}_1)$, and c) using formulas (6)-(8) to obtain $(\mathrm{R}_0, \mathrm{RD})$, $(\mathrm{R}_0, \mathrm{RR})$ and $(\mathrm{R}_0, \mathrm{AP})$.

The 50% and 95 % confidence regions for $(\mathrm{O}_0, \mathrm{OR})$, $(\mathrm{R}_0, \mathrm{RD})$, $(\mathrm{R}_0, \mathrm{RR})$ and $(\mathrm{R}_0, \mathrm{AP})$ are shown in Figures 1-4 respectively.

### 4.3 Confidence intervals of treatment effect alone

Here we are going to calculate the confidence intervals of measures of treatment alone in order to compare them with the confidence regions of measures of (baseline, treatment effect) above. From (5)-(8) or (12)-(15), we see that the measures of treatment effect, OR, RD, RR and AF, are not convertible from one another, except for RR and AF. Hence we need to use distribution of the ML estimate of each measure of treatment effect to obtain $(1-\alpha)$ confidence interval of the measure. We obtain that the 95 % confidence interval of OR is (0.79, 4.78), that for RD is ($-$0.05, 0.36), that for RR is (0.92, 1.92), and that for AF is ($-$0.09, 0.48). These confidence intervals are also presented in Table 3. See Wang et al. (2014a) for interval estimation of the marginal and conditional measures of treatment effect and Wang et al. (2014b) for properties of the ML estimate of RD based on logistic model.

### 4.4 Comparison between confidence regions of (baseline, treatment effect) and confidence intervals of treatment effect

The confidence region of a measure of (baseline, treatment effect) indicates possible values for the measure at the $(1-\alpha)$ confidence level. Statistically, all the confidence regions for ($\mathrm{O}_0$, OR), ($\mathrm{R}_0$, RD), ($\mathrm{R}_0$, RR) and ($\mathrm{R}_0$, AF) are equivalent: each point of a confidence region can be converted to a unique point of another confidence region. Use of a specific confidence region is entirely determined by the specific medical context.

The confidence interval of a measure of treatment effect indicates possible values for the measure at certain $(1-\alpha)$ confidence level regardless of values for baseline. For instance, the confidence interval of RD in Table 3 indicates possible values for RD, i.e. $(-0.05, 0.36)$, at the 95 % confidence level, regardless of values for $\mathrm{R}_0$.

If we interpret the confidence interval of a measure of treatment effect together with a value of a measure of baseline, we may get misleading conclusion. For instance, it is misleading to claim a 95 % confidence interval $(-0.05, 0.36)$ of RD at a baseline risk $\mathrm{R}_0 = 0.68$, because, at the upper limit of the confidence interval, RD + $\mathrm{R}_0$= 0.36 + 0.68 is larger than one.

## 5 Conclusions and discussions

To learn significance of treatment effect to the population, one needs to report both baseline and treatment effect. Because the ML estimate of a measure of baseline is highly correlated with that of a measure of treatment effect, one needs to report baseline and treatment effect jointly. To our knowledge, little is seen in the literature of joint estimation of even one measure of (baseline, treatment effect) by a model in observational studies.

In this article, we express common measures of (baseline, treatment effect) in terms of the conditional risk of outcome given treatments and covariates, which is then expressed by a logistic model. Therefore we can use only one logistic model to control confounding covariates and estimate these measures together. Because the expressions of these measures are smooth bounded functions of the logistic model parameters, the performance of the logistic model in estimating confidence region is determined by the performance of the logistic model in estimating its own model parameters.

Because parameters of a logistic model have good asymptotic normality, we use approximate normal distribution of the ML estimates of the logistic model parameters to generate approximate distributions of the ML estimates of various measures of (baseline, treatment effect) and then use the obtained distributions to calculate confidence regions of these measures. Our method can be easily implemented by any software that generates normal distribution in comparison to high complexity of the normal approximation method and the bootstrap method for estimating confidence regions of measures of (baseline, treatment effect).

To conclude the article, we give two immediate extensions of our approach to estimation of confidence regions for measures of (baseline, treatment effect). First, we can replace the covariate-based logistic model by the propensity score-based logistic model to obtain the confidence regions. Due to a relatively small sample size and relatively few confounding covariates of the illustrative example, we have used the covariate-based logistic model to control the confounding covariates. When sample is large with many covariates, one sometimes uses the propensity score of treatment given covariates to control confounding covariates (Rosenbaum and Rubin, 1983; Rosenbaum, 1995). In this case, we can obtain confidence regions for measures of (baseline, treatment effect) by using the propensity score-based logistic model and considering the propensity score as the sole covariate in the model.

Second, we can use other models than logistic model to obtain confidence regions for measures of (baseline, treatment effect). For instance, sometimes, one finds that some model, e.g. log-linear model, describes the data better than logistic model. In this case, we can obtain confidence regions for measures of (baseline, treatment effect) by replacing the logistic model by the log-linear model in our method.

## Acknowledgement

The authors are grateful to Professor Marta Granström for the data and the relevant medical information. This research received no specific grant from any funding agency in the public, commercial, or not-for-profit sectors.

## References

Austin, P. C. (2010). Absolute risk reductions, relative risks, relative risk reductions, and numbers needed to treat can be obtained from a logistic regression model. *Journal of Clinical Epidemiology* **63** 2--6.

Austin, P. C. (2007). The performance of different propensity score method for estimating marginal odds ratio. *Statistics in Medicine* **26** 3078–3094.

Austin, P. C., Grootendorst, P., Normand, S. L. T. and Anderson, G. M. (2007). Conditioning on the propensity score can result in biased estimation of common measures of treatment effect: A Monte Carlo study. *Statistics in Medicine* **26** 754–768.

Bieler, G. S., Brown, G. G., Williams, R. L. and Brogan, D. J. (2010) Estimating Model-Adjusted Risks, Risk Differences, and Risk Ratios From Complex Survey Data. *American Journal of Epidemiology* **171** 618--623.

Carpenter, J. and Bithell, J. (2000). Bootstrap confidence intervals: when, which, what? A practical guide for medical statisticians. *Statistics in Medicine* **19** 1141—1164.

Cheung, Y. B. (2007). A Modified Least-Squares Regression Approach to the Estimation of Risk Difference. American Journal of Epidemiology 166: 1337-1344.

Davison, A. C. and Hinkley, D. V. (1997). *Bootstrap Methods and their Application*. Cambridge: Cambridge University Press.

Gail, M. H., Wieand, S. and Piantadosi, S. (1984). Biased estimates of treatment effect in randomized experiments with nonlinear regressions and omitted covariates. *Biometrika* **71** 431-- 444.

Gehrmanna, U., Kussb, O., Wellmannc, J. and Ralf Bender, R. (2010). Logistic regression was preferred to estimate risk differences and numbers needed to be exposed adjusted for covariates. *Journal of Clinical Epidemiology* **63** 1223—1231.

Graubard, B. I. and Korn, E. L. (1999). Predictive Margins with Survey Data. *Biometrics* **55** 652--659.

Greenland, S. (2004a). Model-based Estimation of Relative Risk and Other Epidemiologic measures in Studies of Common Outcomes and Case-Control Studies. *American Journal of Epidemiology* **160** 301--305.

Greenland, S. (2004b). Interval estimation by simulation as an alternative to and extension of confidence intervals. *International Journal of Epidemiology* **33** 1389–1397

Greenland, S. (1987). Interpretation and choice of effect measures in epidemiologic analyses. *American Journal of Epidemiology* **125** 761--768.

Greenland, S., Robins, J. M. and Pearl, J. (1999). Confounding and collapsibility in causal inference. *Statistical Science* **14** 29–46.

Greenland, S. and Robins, J. M. (1986). Identifiability, exchangeability, epidemiological confounding. *International Journal of Epidemiology* **15** 413–419.

Lane, P. W. and Nelder, J. A. (1982). Analysis of Covariance and Standardization as Instance of Prediction. *Biometrics* **36** 613--621.

Lee, Y. and Nelder, J. A. (2004). Conditional and marginal models: another view. *Statistical Science* **19** 219--228.

Lindsey, J. K. (1996). *Parametric Statistical Inference.* Oxford: Clarendon Press.

McNutt, L. A., Wu, C., Xue, X. and Hafner, J. P. (2003). Estimating the relative risk in cohort studies and clinical trials of common outcomes. *American Journal of Epidemiology* **157** 940--943.

Nguyen, T. V. H., Bengtsson, C., Nguyen, G. K., Hoang, T. H., Phung, D. C., Sörberg, M. and Granström, M. (2008). Evaluation of Two Triple-Therapy Regiments with Metronidazole or Clarithromycin for the Eradication of *H. pylori* Infection in Vietnamese Children, a Randomized Double Blind Clinical Trial. *Helicobacter* **13** 550—556.

Rosenbaum, P. R. (1995). *Observational studies*. New York: Springer.

Rosenbaum, P. R. and Rubin, D. B. (1983). The central role of the propensity score in observational studies for causal effects. *Biometrika* **70** 41--55.

Rubin, D. B., Wang, X., Yin, L., and Zell, E. (2009), “Estimating the Effect of Treating Hospital Type on Cancer Survival in Sweden Using Principal Stratification” in The HANDBOOK OF APPLIED BAYESIAN ANALYSIS, eds. T. O’Hagan and M. West, Oxford University Press, Oxford.

Spiegelman, D. and Hertzmark, E. (2005). Easy SAS calculations for risk or prevalence ratios and differences. *American Journal of Epidemiology* **162** 199--200.

Wang, X., Jin, Y. and Yin, L. (2014a). Measuring and estimating treatment effect on dichotomous outcome of a population. *Statistical methods in medical research* doi:10.1177/0962280213502146.

Wang, X., Jin, Y. and Yin, L. (2014b). Point and interval estimations of marginal risk difference by logistic model. *Communications in statistics - Theory and Methods*, to appear.

Zhang, J., Yu, K. F. (1998). What’s the relative risk? A method of correcting the odds ratio in cohort studies of common outcomes. *JAMA* **280** 1690–1691.

Zou, Y. (2014). A Modified Poisson Regression Approach to Prospective Studies with Binary Data. *American Journal of Epidemiology* **159** 702--706.

**Table 1** Eradication of *Helicobacter pylori* among Vietnamese children: successful eradications / total patients on levels of the covariates $x_1$ -- $x_4$ for high respective low doses of the treatment $z$.

| **Covariates** | | | | **Treatment** | |
|---|---|---|---|---|---|
| $x_1$ (age) | $x_2$ (gender) | $x_3$ (geographic) | $x_4$ (resistance) | $z = 1$(high) | $z = 0$(low) |
| 0 | 0 | 0 | 0 | 0/3 | 3/4 |
| 0 | 0 | 0 | 1 | 1/1 | 3/4 |
| 0 | 0 | 1 | 0 | 8/8 | 6/7 |
| 0 | 0 | 1 | 1 | 1/1 | |
| 0 | 1 | 0 | 0 | 2/3 | 1/5 |
| 0 | 1 | 0 | 1 | 1/2 | 2/4 |
| 0 | 1 | 1 | 0 | 3/3 | 3/4 |
| 0 | 1 | 1 | 1 | | 1/1 |
| 1 | 0 | 0 | 0 | 3/8 | 1/2 |
| 1 | 0 | 0 | 1 | 1/1 | 3/3 |
| 1 | 0 | 1 | 0 | 6/6 | 1/1 |
| 1 | 0 | 1 | 1 | 3/3 | 0/1 |
| 1 | 1 | 0 | 0 | 5/8 | 2/7 |
| 1 | 1 | 0 | 1 | 6/9 | 1/2 |
| 1 | 1 | 1 | 0 | 0/2 | 0/2 |
| 1 | 1 | 1 | 1 | 1/1 | 0/3 |

**Table 2** ML estimates and their approximate covariance matrix for parameters of model (1)
.

| Parameters | $\alpha$ | $\beta$ | $\theta_1$ | $\theta_2$ | $\theta_3$ | $\theta_4$ | $\theta_5$ | $\theta_6$ |
|---|---|---|---|---|---|---|---|---|
| Estimates | 1.19 | −0.87 | −0.57 | −1.82 | 0.10 | 0.55 | 1.96 | 2.18 |
| Covariance matrix | | | | | | | | |
| $\alpha$ | 0.64 | −0.53 | −0.12 | −0.42 | −0.32 | −0.13 | 0.47 | 0.29 |
| $\beta$ | −0.53 | 1.06 | −0.14 | 0.45 | 0.31 | 0.10 | −0.89 | −0.67 |
| $\theta_1$ | −0.12 | −0.14 | 0.37 | −0.04 | 0.00 | −0.05 | 0.03 | 0.05 |
| $\theta_2$ | −0.42 | 0.45 | −0.04 | 0.69 | 0.05 | −0.01 | −0.69 | −0.05 |
| $\theta_3$ | −0.32 | 0.31 | 0.00 | 0.05 | 0.72 | 0.06 | −0.07 | −0.71 |
| $\theta_4$ | −0.13 | 0.10 | −0.05 | −0.01 | 0.06 | 0.39 | −0.10 | −0.04 |
| $\theta_5$ | 0.47 | −0.89 | 0.03 | −0.69 | −0.07 | −0.10 | 1.40 | 0.32 |
| $\theta_6$ | 0.29 | −0.67 | 0.05 | −0.05 | −0.71 | −0.04 | 0.32 | 1.86 |

**Table 3** ML estimates and confidence intervals for measures of baseline and treatment effect based on model (1).

| | ML estimate | 50 % and 95 % Confidence intervals |
|---|---|---|
| Baseline measure | | |
| $R_0$ | 0.53 | (0.37, 0.48, 0.58, 0.68) |
| $O_0$ | 1.13 | (0.58, 0.92, 1.38, 2.13) |
| Effect measure | | |
| OR | 2.11 | (0.79, 1.45, 2.67, 4.78) |
| RR | 1.33 | ( 0.92, 1.15, 1.47, 1.92) |
| RD | 0.17 | (−0.05, 0.09, 0.23, 0.36) |
| AF | 0.25 | (−0.09, 0.13, 0.32, 0.48) |

**Figure 1** (a) The scatterplot for approximate distribution of the ML estimate $(\widehat{O}_0, \widehat{OR})$, and the 50% and 95% confidence regions of $(O_0, OR)$. (b) Approximate distribution for the ML estimate $\widehat{O}_0$. (c) Approximate distribution for the ML estimate $\widehat{OR}$. The point estimate of $(O_0, OR)$ is equal to $(1.13, 2.11)$.

(a)

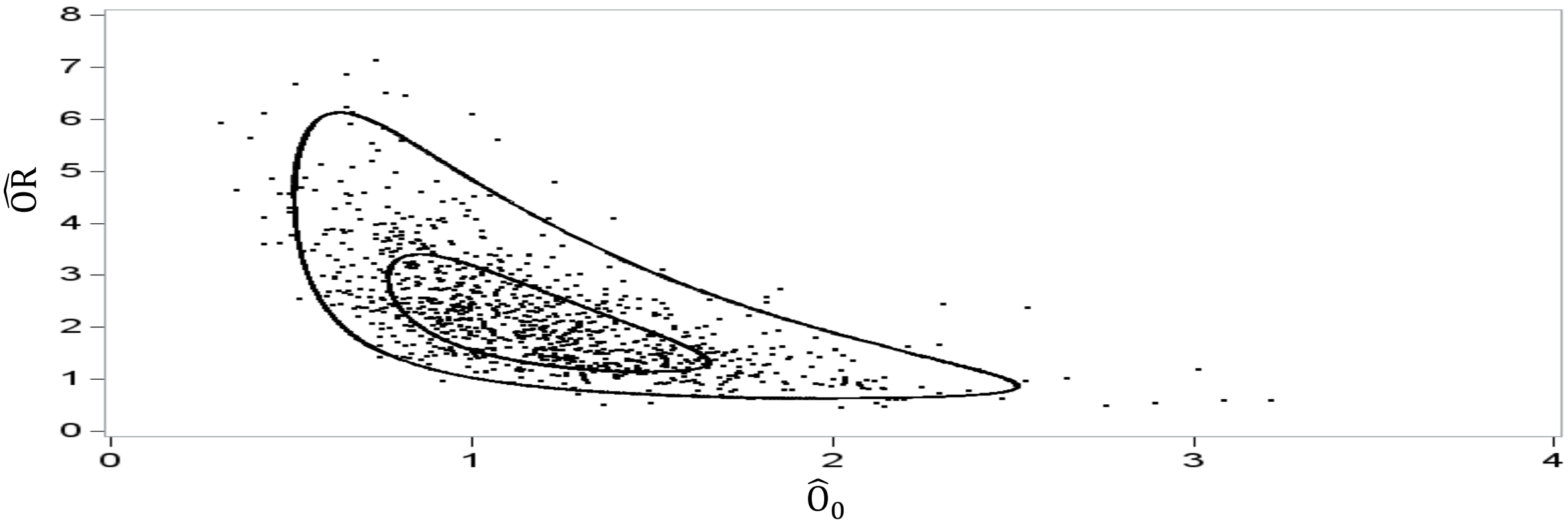

(b)

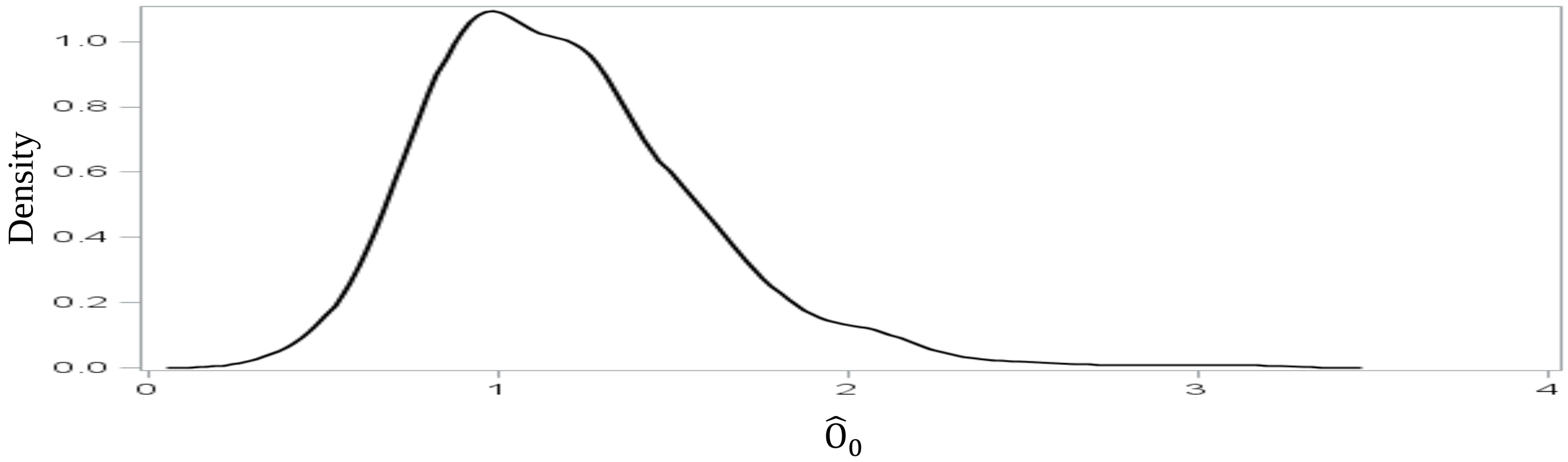

(c)

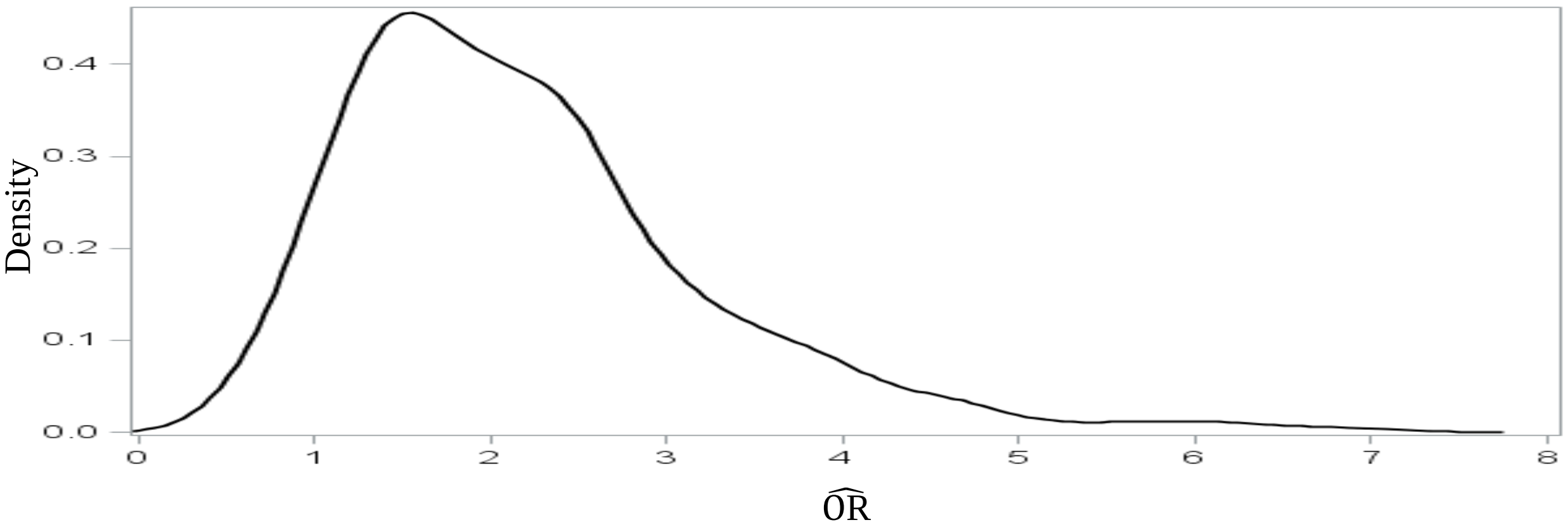

**Figure 2** (a) The scatterplot for approximate distribution of the ML estimate $(\widehat{R}_0, \widehat{RD})$, and the 50% and 95% confidence regions of $(R_0, RD)$. (b) Approximate distribution for the ML estimate $\widehat{R}_0$. (c) Approximate distribution for the ML estimate $\widehat{RD}$. The point estimate of $(R_0, RD)$ is equal to $(0.53, 0.17)$.

(a)

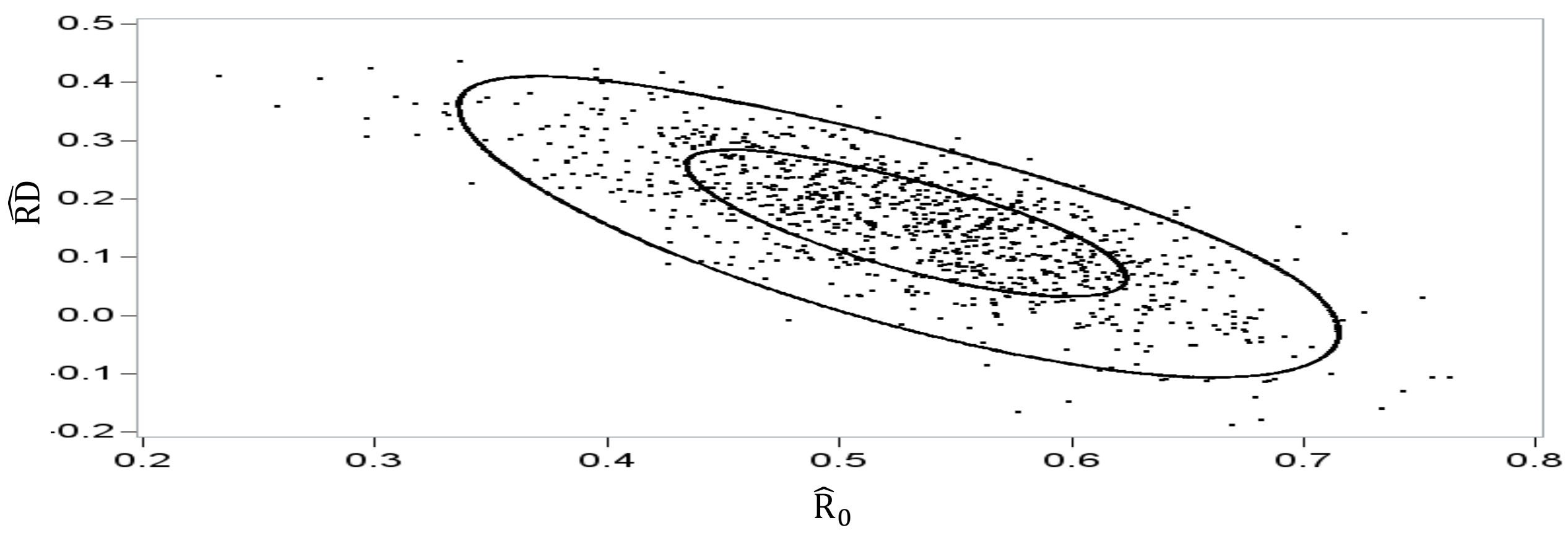

(b)

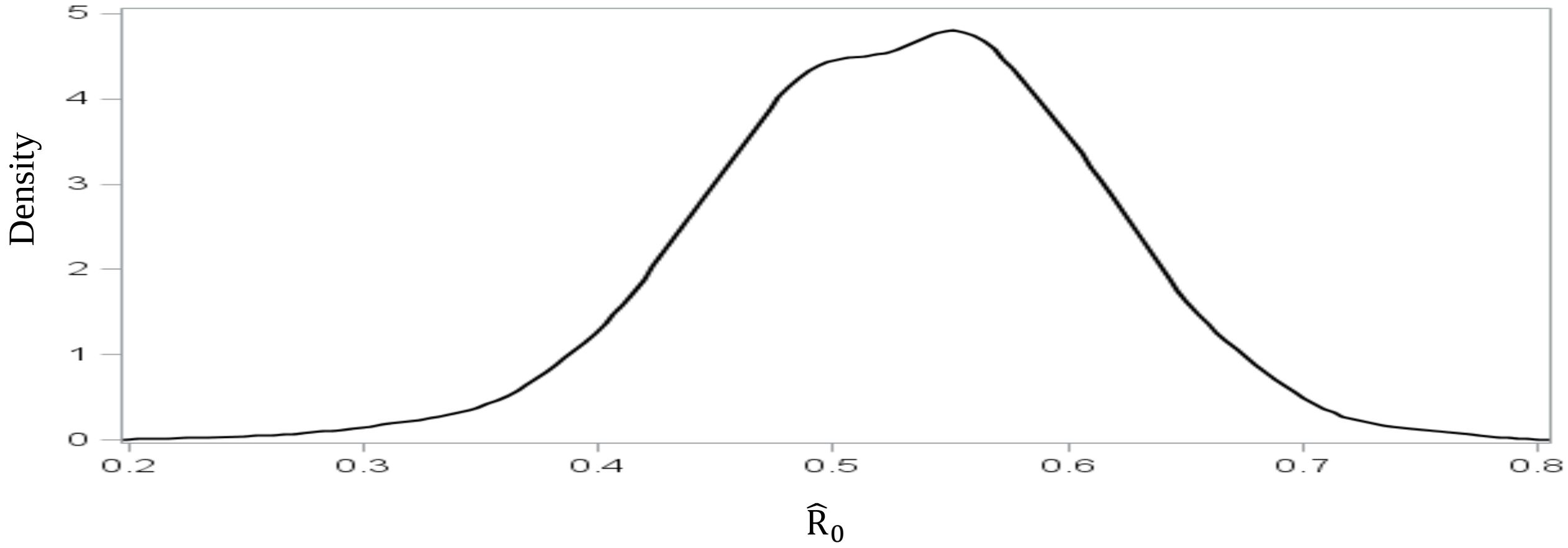

(c)

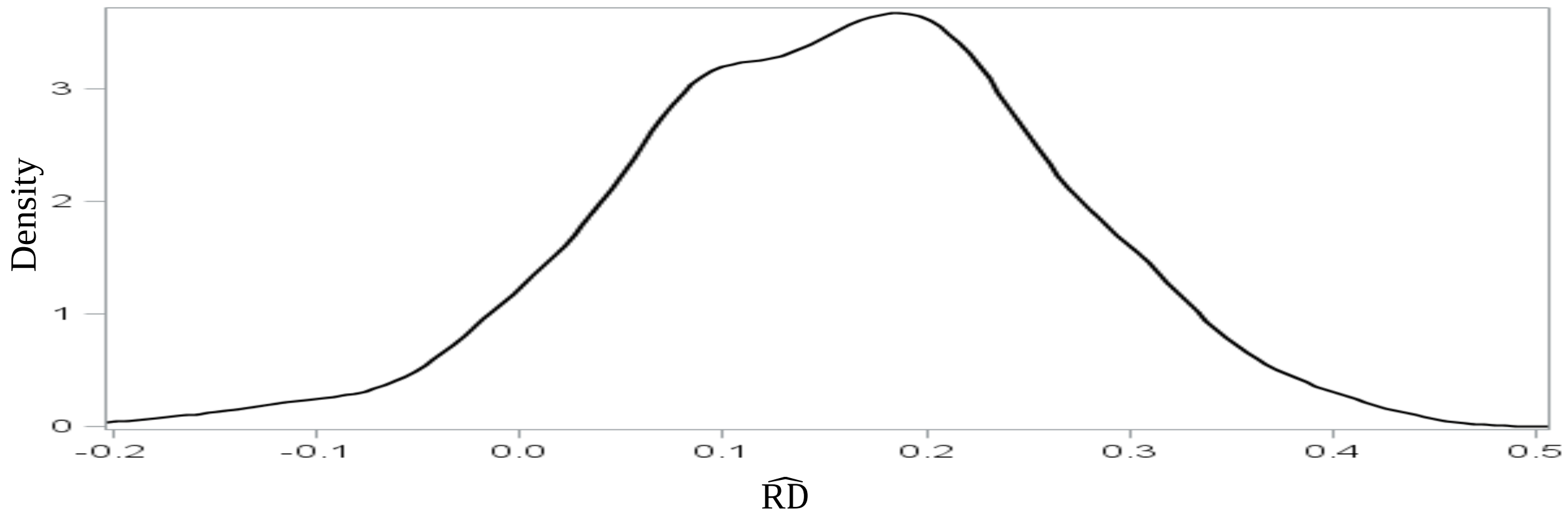

**Figure 3** (a) The scatterplot for approximate distribution of the ML estimate $(\widehat{R}_0, \widehat{RR})$, and the 50% and 95% confidence regions of $(R_0, RR)$. (b) Approximate distribution for the ML estimate $\widehat{RR}$. The point estimate of $(R_0, RR)$ is equal to $(0.53, 1.33)$.

(a)

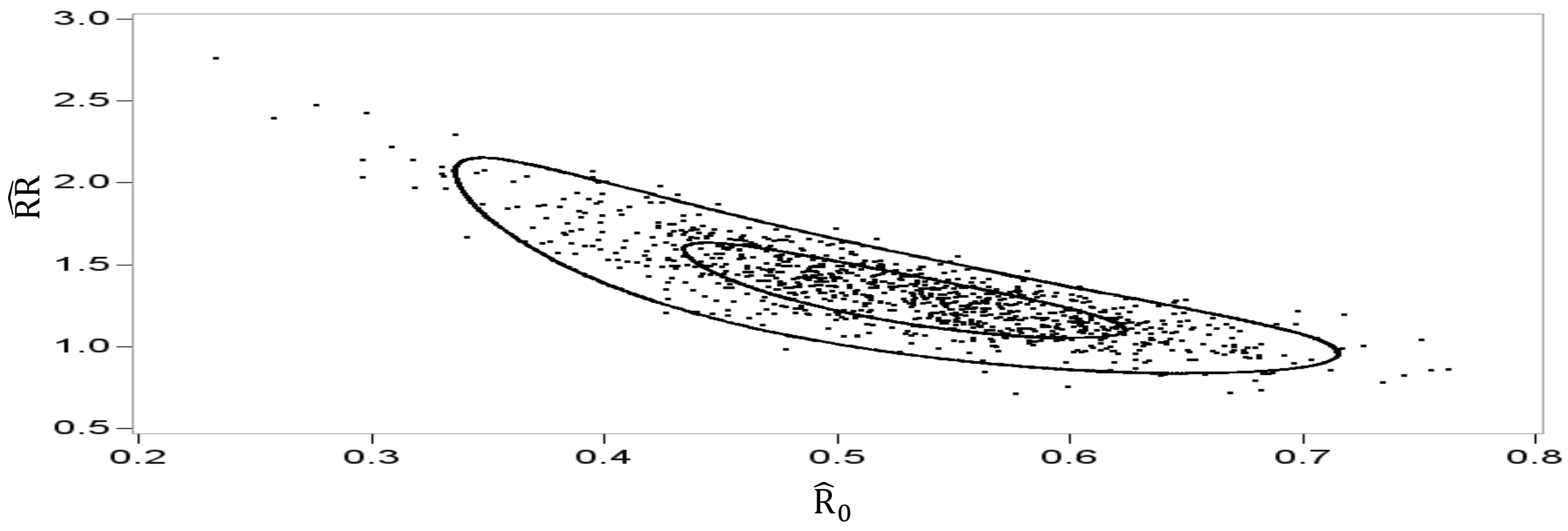

(b)

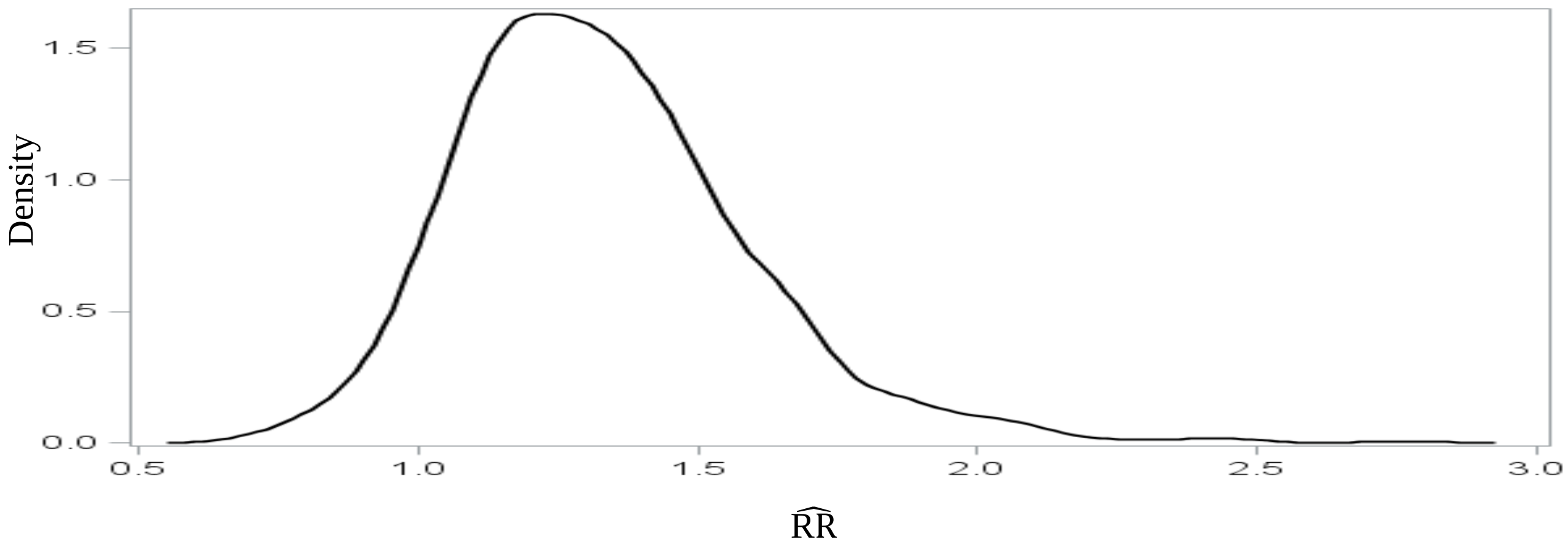

**Figure 4** (a) The scatterplot for approximate distribution of the ML estimate $(\widehat{R}_0, \widehat{AF})$, and the 50% and 95% confidence regions of $(R_0, AF)$. (b) Approximate distribution for the ML estimate $\widehat{AF}$. The point estimate of $(R_0, AF)$ is equal to $(0.53, 0.25)$.

(a)

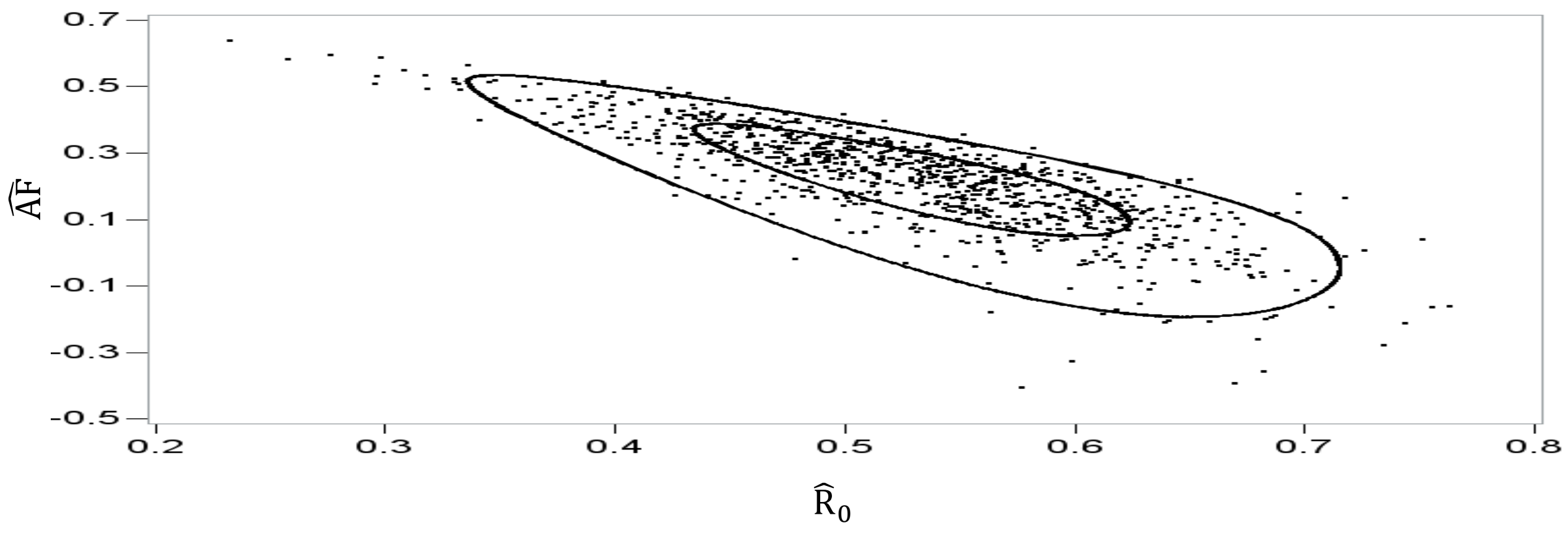

(b)

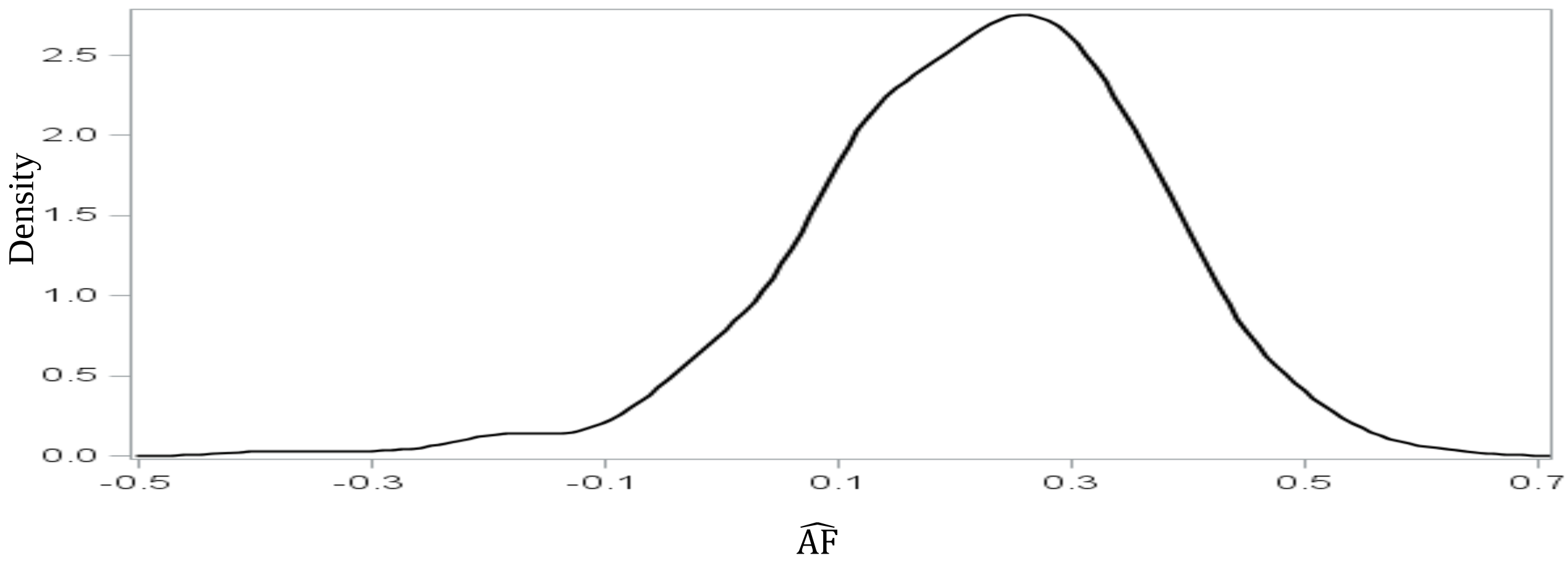

**Figure 5** (a) The scatterplot for approximate distribution of the ML estimate $\{\log(\widehat{\mathrm{O}}_0)$, $\log(\widehat{\mathrm{OR}})\}$, and the 50% and 95% confidence regions of $\{\log(\mathrm{O}_0), \log(\mathrm{OR})\}$. (b) Approximate distribution for the ML estimate $\log(\widehat{\mathrm{O}}_0)$. (c) Approximate distribution for the ML estimate $\log(\widehat{\mathrm{OR}})$. The point estimate of $\{\log(\mathrm{O}_0), \log(\mathrm{OR})\}$ is equal to $(0.12, 0.75)$.

(a)

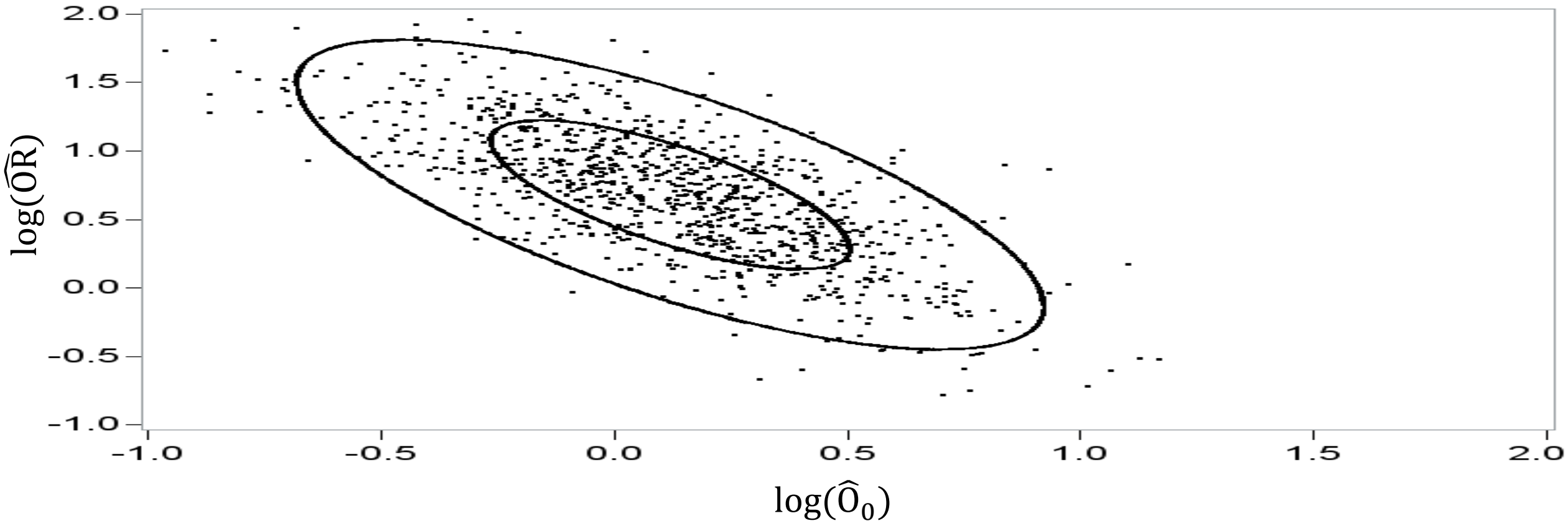

(b)

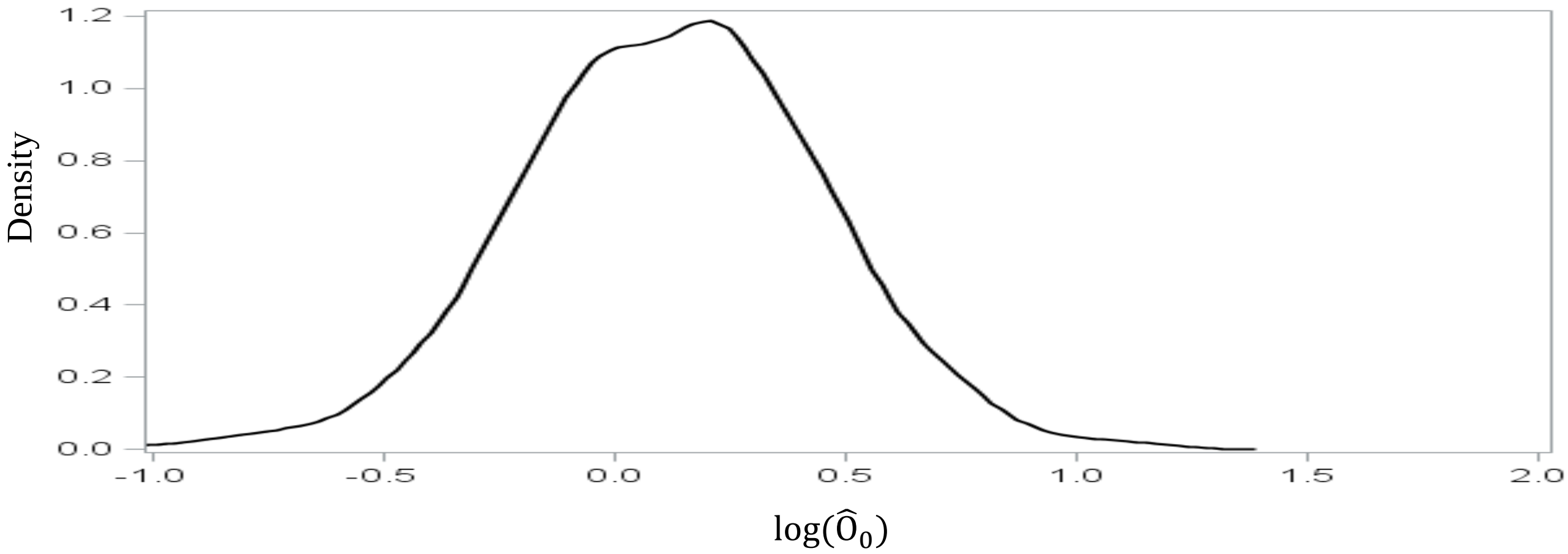

(c)

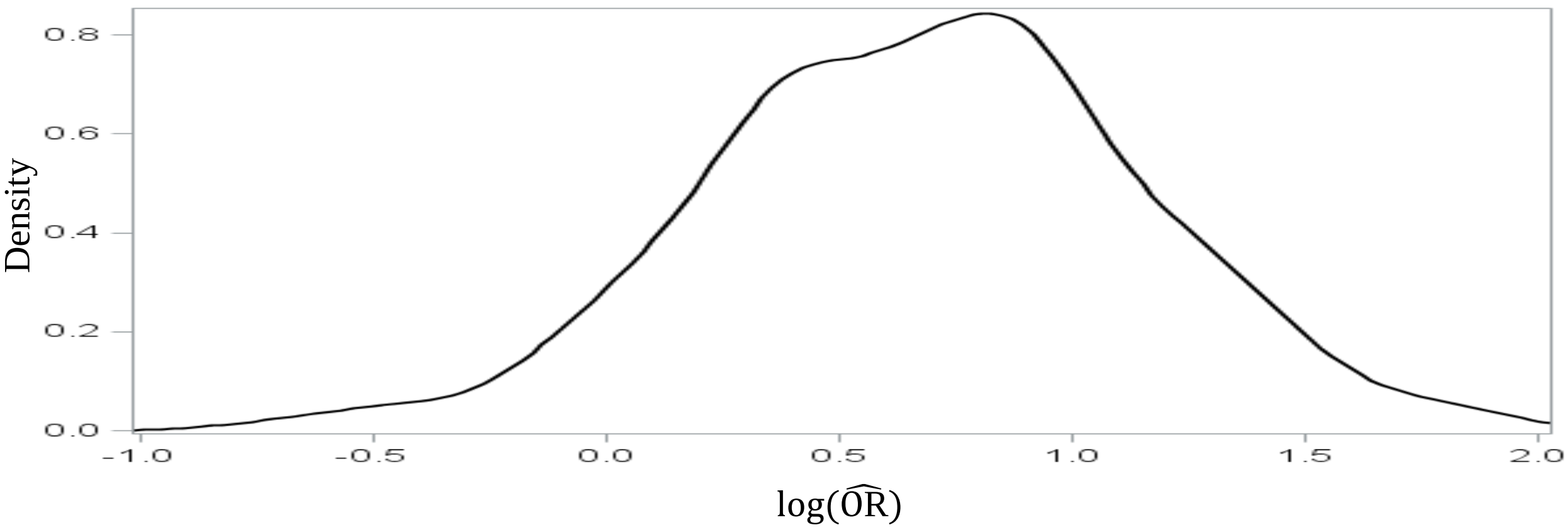